\documentclass[useAMS, usenatbib]{mn2e}
\usepackage{times}
\usepackage{epsfig}
\usepackage{multirow}
\usepackage{natbib}
\usepackage{amsmath}
\usepackage{psfrag}
\usepackage{color}
\usepackage[psamsfonts]{amssymb}

\newcommand {\remove}[1]{}

\newcommand {\beq}{\begin{equation}}
\newcommand {\eeq}{\end{equation}}
\newcommand {\eq}[1]{eq.~(\ref{#1})}
\newcommand {\msun}{{\rm M}_{\odot}}

\def\1915{GRS 1915+105}

\title{The outburst duration and duty-cycle of GRS\,1915+105} 
\author[P. Deegan, C. Combet and G. Wynn]{Patrick
  Deegan\thanks{E-mail: pd75@astro.le.ac.uk}, C\'eline Combet and Graham A. Wynn \\
   Department of Physics and Astronomy, University of Leicester, Leicester, LE1
  7RH, United Kingdom}

\date{Accepted ??. Received ??; in original form \today}

\volume{000}

\begin{document}

\label{firstpage}

\maketitle

\begin{abstract}
The extraordinarily long outburst of \1915 makes it one of the most
remarkable low-mass X-ray binaries (LMXBs). It has been in a state of
constant outburst since its discovery in 1992, an eruption which has
persisted $\sim$ 100 times longer than those of more typical
LXMBs. The long orbital period of \1915 implies that it contains large
and massive accretion disc which is able to fuel its extreme outburst.
In this paper, we address the longevity of the outburst and quiescence
phases of \1915 using Smooth Particle Hydrodynamics (SPH)
simulations of its accretion disc through many outburst cycles. Our
model is set in the two-$\alpha$ framework and includes the effects of 
the thermo-viscous instability, tidal torques, irradiation by central
X-rays and wind mass loss.  We explore the model parameter space and
the examine the impact of the various ingredients.  We predict that
the outburst of \1915 should last a minimum of 20 years and possibly
up to $\sim 100$ years if X-ray irradiation is very significant. The
predicted recurrence times are of the order of $10^{4}$ years, making
the X-ray duty cycle a few $0.1\%$. Such a low duty cycle may mean
that \1915 is not an anomaly among the more standard LMXBs and that
many similar, but quiescent, systems could be present in the Galaxy.
\end{abstract}


\section{Introduction}\label{sec:introduction}
Low mass X-ray binaries (LMXB) are bright X-ray sources consisting of
an accreting black hole/neutron star primary and a Roche-lobe filling,
low-mass secondary star.  Matter is transferred from the secondary
star via the inner-Lagrange (L1) point and forms an accretion disc
around the primary. In some cases the accretion disc undergoes
sporadic outbursts which are thought to be triggered by a
thermal-viscous instability, resulting in an increased mass accretion
rate onto the primary (e.g. \citealp{lasota01}) and
a rapid X-ray brightening. Many transient systems have been observed in 
the local neighbourhood (Milky Way, LMC, SMC, e.g. \citet{2007A&A...469..807L}) and
the duration of the outbursts is typically of the order of months.  The
quiescence period between outbursts in these systems generally lasts one to a
few years, making their duty cycle $\sim 1\%$.
Such a value for the duty cycle, based only on these short period well-studied 
systems, is the one generally assumed in the models of X-ray luminosity functions of nearby 
galaxies (e.g. \citealp{2004ApJ...601L.147B}). However, nothing suggests that it holds for 
the systems that have not been observed going through a full cycle. This is the case, for example, 
of some long-duration transients for which the quiescence phase duration, hence duty cycle, 
can only be indirectly estimated from the cooling curve of the neutron star 
primary\footnote{For example, \citet{2001ApJ...560L.159W} estimated that the quiescence
phase of the quasi-persistent transient KS 1731-260 could be several hundreds years long.}
(e.g. \citealp{2001ApJ...560L.159W, 2008ApJ...687L..87C}). Furthermore, such a
value of the duty cycle could be valid for the shorter period systems only but not for their 
longer period counterparts. It is this latter question we address in this work by exploring 
the case of the long period system GRS 1915.

\1915 is one of the most well studied LMXBs (see \citet{Fender04}
for a review). Recent observations \citep{Harlaftis04}
suggest the system contains a 14.0 $\pm$ 4.4 $\msun$ black hole with
a 0.8 $\pm$ 0.5 $\msun$ secondary, in a binary of orbital period 33.5
days \citep{Greiner01b}.  The outburst which lead to the system's
detection in 1992 \citep{Castro92} is still proceeding to this day,
lasting 20 times longer than any other LMXB. This longevity can be
explained in simple terms by the long orbital period, which means
there is plenty of room in the system for a very large accretion disc
to form around the black hole ($R_{\rm disc} \sim 2 \times 10^{12}$~cm,
\citealp{Truss06}). A large disc ensures there is a large reservoir of
mass available to fuel the long outburst.  Analytic estimates of the
duration of \1915's outburst were derived by \cite{Truss06}. By their
nature, these estimates were not able to account for the dynamical and
viscous evolution of the accretion disc. In this paper we re-address
the duration of the outbursts of \1915 and also consider its X-ray
duty cycle using smooth particle hydrodynamics (SPH) simulations of the
accretion disc through many outburst cycles. In section
\ref{sec:ingredients} we present the physical effects included in the
simulations and the associated free
parameters. Section~\ref{sec:numerics} deals with the implementation
of these ingredients in the SPH
code.  In section \ref{sec:results} we present our results and
estimate the duty cycle of \1915. In the last section we discuss the
implications of this work.

\section{Physical ingredients}
\label{sec:ingredients}

\subsection{The disc instability model (DIM) \label{subsec:dim}}
The disc instability model was first 
developed to explain the outbursts of cataclysmic variables (dwarf
novae) and has been successfully extended to explain the 
X-ray outbursts of LMXBs \citep{1989ApJ...343..241M,dubus01}. 
Here, we briefly address the DIM and refer the reader to \citet{lasota01} for a review.

Vertical (the direction perpendicular to the orbital plane of the binary)
thermal equilibrium in the disc results in a relation between the 
surface density of the disc $\Sigma$ and its temperature $T$ at any given 
radius.  The locus of equilibrium positions in the $\Sigma-T$~plane,
takes the shape of an \emph{S-curve}.
The negatively sloped middle branch is caused by
the sudden change of opacity as hydrogen ionises at $T\gtrsim
6500$~K. When conditions dictate that the equilibrium 
position is situated on this middle branch the disc is forced to  
follow a limit-cycle between hot and
cold states, defined by  
the critical surface densities $\Sigma _{\rm max}$ and $\Sigma _{\rm
min}$. These hot and cold states are thought to be respectively associated with 
enhanced and suppressed angular momentum transport within the disc. 
Many solutions utilise  the
Shakura-Sunyaev viscosity prescription \citep{shakura73}, namely, 
\beq\label{eqn:SS_vis}
\nu=\alpha c_s H\;, 
\eeq 
where $H$ is the scale height, $c_{s}$ is the
local sound speed in the disc, and $\alpha$ a free parameter. 
The hot state is assumed to be associated with high disc viscosity $\alpha_h$, and 
the cold state with low disc viscosity $\alpha_c$. This two-alpha approach was
introduced early in the development of the models to reproduce observed outbursts durations. 
This translates into $\Sigma_{\rm max}=\Sigma_{\rm
max}(\alpha_c)$ and $\Sigma_{\rm min}=\Sigma_{\rm min}(\alpha_h)$. If
anywhere in the disc $\Sigma(r)> \Sigma_{\rm max}$, the annulus enters
the hot, high-viscous state, which propagate to nearby annuli. The
front propagating inward forces the disc in the hot viscous state on
its way: this high viscosity implies a high accretion rate onto the
central object, leading to the X-ray outburst. The disc returns to
quiescence (low viscosity, small accretion rate) once $\Sigma(r)<
\Sigma_{\rm min}$.

It has been found that $\Sigma_{\rm max}$ and
  $\Sigma_{\rm min}$ scale (almost) linearly with radius \citep{cannizzo88}:
\begin{equation}
\Sigma_{\rm max} = 11.4 R_{10}^{1.05} M_{1}^{-0.35} \alpha_c^{-0.86}
{\rm \,g\,cm}^{-2} \;,
\label{eqn:sigmamax}
\end{equation}
and   
\begin{equation}
\Sigma_{\rm min} = 8.25 R_{10}^{1.05} M_{1}^{-0.35}\alpha_h^{-0.8}
{\rm \,g\,cm}^{-2} \;,
\label{eqn:sigmamin}
\end{equation}
where $M_1$ is the primary mass in solar masses, and $R_{10}$ is the
radius in units of $10^{10}$~cm. Equations (\ref{eqn:sigmamax}) and
(\ref{eqn:sigmamin}) are used in our numerical setup, but note that some
slightly different prescriptions exist.


\subsection{Irradiation \label{subsec:irradiation}}
As discussed above wherever $\Sigma(r) > \Sigma_{\rm max}$, the disc is in the hot
highly-viscous state. However, this is not the only grounds for the disc entering the
hot state. X-ray irradiation by the central object can keep the disc ionised
and in the hot state out to a radius $R_{\rm irr}$. For
typical black hole and disc parameters, \citet{king1998} find
\begin{equation}
 R_{\rm irr} \sim 2.7 \times 10^{11}
\left(\frac{\eta}{0.1}\right)^{1/2}\left(\frac{\epsilon}{10^{-3}}\right)^{1/2}\dot{M}_{18}^{1/2}
{\rm cm},
\label{eqn:irr}
\end{equation}
 where $\dot{M}_{18}$ is the central accretion rate in units of $\rm
   10^{18}$~g~s$^{-1}$, $\eta$ the accretion efficiency (we keep $\eta=0.1$
   throughout this work)
   and $\epsilon$ the illumination efficiency. The latter parametrises the
   geometric properties of the disc (incident angle of irradiation, albedo),
   whereas $\eta$ gives the fraction of the accretion energy that is radiated
   away.  The accretion onto the black hole can, at most, proceed at the
   Eddington rate,
\beq
\dot{M}_{\rm Edd}^0 = \frac{R_0 L_{\rm Edd}}{G M_{1}}=3.1\times
10^{-7}\;\msun\;{\rm yr}^{-1} 
\eeq
for \1915 ($R_0$ is the black hole radius). Hence, from \eq{eqn:irr}, one gets the largest extent of the disc
that can be switched to the hot state {\em via} accretion-powered radiation is
\beq
R_{\rm Edd} = 1.21 \times 10^{12} \left(\frac{\epsilon}{10^{-3}} \right)^{1/2}{\rm ~cm}.
\label{eqn:redd}
\eeq
We cap the irradiation radius at its Eddington value so that $R_{\rm
  irr}=R_{\rm Edd}$ whenever $\dot M_1 > \dot M_{\rm Edd}^0$.

\subsection{Wind Loss \label{wind}}
Mass loss due to the local mass transfer rate exceeding the Eddington
limit is also included in the model. The local accretion rate at radius $r$ is given
by,
\begin{equation}
\dot{M}(r) = -2\pi r v_{r}(r) \Sigma (r)
\end{equation}
where $v_r(r)$ is the radial velocity. The Eddington rate at the same radius
is,
\begin{equation}
\dot{M}_{\rm Edd}(r) = \frac{r L_{\rm Edd}}{G M_{1}},
\label{eq:edd_rate}
\end{equation}
where $L_{\rm Edd}$ is the Eddington luminosity. 
If $\dot{M}(r) > \dot{M}_{\rm Edd}(r)$ a wind should carry away the
energy in excess of the Eddington limit. To allow for some freedom in the 
wind efficiency, we parametrise the latter by triggering the wind whenever
\beq
\dot{M}(r) > \lambda \dot{M}_{\rm Edd}(r)\;.
\label{eqn:wind_threshold}
\eeq
The parameter $\lambda$ allows us to depart from the strict Eddington limit ($\lambda=1$): the smaller
$\lambda$, the smaller the threshold at which a wind is emitted. 

The three mechanisms detailed above contain the free
parameters that are varied throughout this work in order to reveal their
relative effects, namely: $\alpha_h$, $\alpha_c$, $\epsilon$
and $\lambda$. Note also that the expressions for $\Sigma_{\rm max}$
and $\Sigma_{\rm min}$ are estimates only and that we allow for a change in
their normalisations which gives us two extra parameters. This is detailed in
the numerical setup below.

\section{Numerical implementation} 
\label{sec:numerics}
We study the long term evolution of \1915 using the SPH 
code first developed by \citet{murray1996} and later
modified by \citet{Truss04}. It includes the
thermal viscous instability, disc irradiation, and wind loss as described
above. Moreover, the system is followed in the full binary potential, hence it
naturally includes any tidal effects that may arise.

\subsection{Particle injection and rejection \label{subsec:par_inj_rej}}
The particles are injected from the L1 point, and into the
primary's potential, with the transfer rate given by \citet{Ritter99}:
\begin{equation}
 -\dot{M}_2 \sim 7.3 \times 10^{-10} \left( \frac{M_2}{\msun}\right)^{1.74}
\left(\frac{P_{\rm orb}}{{\rm 1\;day}}\right)^{0.98}\;\msun \;{\rm yr}^{-1}\;.
\label{masstran}
\end{equation}
Using \1915 system characteristics, this gives $-\dot M_2\sim 2\times
10^{-8}\msun$~yr$^{-1}$.

Particles are removed from the simulation when they are within 0.04$a$ ($a$ is
the binary separation) of the black hole, if they return to the secondary's
Roche lobe or if they are at a distance $r>a$ with a velocity greater than the 
escape velocity. The first condition
implies that the accretion rates we derive are not the accretion rates onto
the black hole (as $0.04a\gg R_{\rm Schw}$ where $R_{\rm Schw}$ is the
Schwarzschild radius of the black hole). Some material may still be blown
away in a wind before it reaches the black hole surface but resolution issues prevent
us from studying the most inner regions of the accretion disc.

\subsection{Triggering the disc instability \label{subsec:inst_num}}
Practically, the disc is divided into one hundred annuli into which
the surface density is evaluated. Whenever
$\Sigma(r)>\Sigma_{\rm max}(r)$, the corresponding ring is switched into the hot state
($\alpha=\alpha_h$).  Conversely, the disc switches back to the cold
state ($\alpha=\alpha_c$) wherever $ \Sigma<\Sigma_{\rm min}$. Using
Eqs.~(\ref{eqn:sigmamax}) and (\ref{eqn:sigmamin}) with the parameters of
\1915 and the typical values $\alpha_h=0.1$ and $\alpha_c=0.01$ , one gets

\begin{equation}
\Sigma_{\rm max} = K_{\rm max}\left(\frac{r}{a}\right)^{1.05}\sim 2.5\times
10^{5}\left(\frac{r}{a}\right)^{1.05} {\rm \,g\,cm}^{-2}\,
\label{eqn:gradmax}
\end{equation}
and
\begin{equation}
\Sigma_{\rm min} = K_{\rm min}\left(\frac{r}{a}\right)^{1.05}\sim 2.2\times
10^{4}\left(\frac{r}{a}\right)^{1.05}{\rm \,g\,cm}^{-2} \,.
\label{eqn:gradmin}
\end{equation}
In practice, we cannot use the values of $ K_{\rm max/min}$ shown
above. 
Building the disc up to $K_{\rm max}\sim 2.5\times 10^{5} {\rm
\,g\,cm}^{-2}$ would take a prohibitive amount
of time: $K_{\rm max}^{\rm sph}=55$ and $K_{\rm min}^{\rm sph}=10$
are typical values we use. The quantities $K_{\rm max}$ and $K_{\rm min}$
are also to be varied to explore how a less or more massive disc
would behave.

Along the same lines, $\alpha_h$ and $\alpha_c$ are also increased
from their canonical values given above, in order to speed up the outburst and
recurrence times. Doing so allows the system to quickly reach
steady-state and undergo several outburst events during a single run. The
draw-back is that results need to be scaled in order to get actual the
recurrence and outburst times. This is detailed in \S\ref{subsec:scalings}
below.

The last technical point is that of the disc transition between the
cold and hot states. It is performed following \citet{Truss04}, and is set up
to occur on the thermal time scale of the system, $t_{\rm th}\sim
\Omega_K^{-1}$ ($\Omega_K$ is the keplerian frequency).  Once a switch is
triggered, $\alpha$ follows
\begin{equation}\label{eqn:alpha_t}
\alpha(t) = \alpha^{+} \pm \alpha^{-} \tanh\left(\frac{t}{t_{\rm
    th}}-\pi\right) \; ,
\end{equation}
where
\beq
\alpha^{\pm} = \frac{(\alpha_h \pm \alpha_c)}{2}\;.
\eeq

\subsection{Scaling outburst and recurrence timescales \label{subsec:scalings}} 
As discussed in \S\ref{subsec:inst_num}, computational time considerations have
forced us to use values for $\Sigma_{\rm max/min}$ and $\alpha_ {h/c}$
which are different from their generally inferred values. In
this section we detail the procedure we follow to scale the raw outburst and
recurrence durations to that of the physical system.  In the
following, we denote by {\tt real} and {\tt sph} the quantities related to the
physical and the simulated system respectively.

During an outburst most of the mass inside a certain radius, $R_{\rm out}$,
will be accreted.  At the start of the outburst, the surface density inside
this radius will be near $\Sigma_{\rm max}$ (see Fig.\ref{fig:disc_evol}, left
panel). The outburst timescale is then given by
\begin{equation}
t_{\rm out}\sim\frac{m}{\dot{m}}\;,
\end{equation}
where $\dot{m}$ is 
the accretion rate in the disc for $r<R_{\rm out}$ and $m$ is the
total mass in that region. The latter is  given by,
\begin{equation}\label{eqn:mass}
m \sim \int^{R_{\rm out}}_{0} 2 \pi r \Sigma_{\rm max}(r) dr -
\int^{R_{\rm out}}_{0} 2 \pi r \Sigma_{\rm min}(r) dr.
\end{equation} 
Using eqns. (\ref{eqn:sigmamax}) and (\ref{eqn:sigmamin}), one gets
\begin{equation}\label{eqn:re_mass}
m \sim \frac{2 \pi R_{\rm out}^2}{3.05}\left[\Sigma_{\rm max}(R_{\rm
out})-\Sigma_{\rm min}(R_{\rm out})\right].
\end{equation} 
Assuming $R_{\rm out}$ is constant, eqns. (\ref{eqn:gradmax}),
(\ref{eqn:gradmin}) and (\ref{eqn:re_mass}) give,
\begin{equation}\label{eqn:mass_k}
m \propto (K_{\rm max} - K_{\rm min})\;.
\end{equation}
Given that (see \citet{Frank02})
\begin{equation}
\dot{m} \propto \alpha_h\Sigma_{\rm max}\propto \alpha_hK_{\rm
  max}
\end{equation}  
then
\begin{equation}
t_{\rm out} \propto \frac{(K_{\rm max} - K_{\rm min})}{\alpha_hK_{\rm
max}}\;.
\end{equation} 
This provides a scaling for the results from the SPH code. The
ratio of the physical outburst time to the simulated one is therefore,
\begin{equation}
\frac{t_{\rm out}^{\rm real}}{t_{\rm out}^{\rm sph}} = \frac{(K_{\rm max}^{\rm
  real} - K_{\rm min}^{\rm real})\alpha_h^{\rm sph}K_{\rm max}^{\rm
  sph}}{(K_{\rm max}^{\rm sph} -
  K_{\rm min}^{\rm sph})\alpha_h^{\rm real}K_{\rm max}^{\rm real}}.
\end{equation}
The quiescence time is scaled in a similar fashion; the time spent in quiescence is roughly the time needed to replenish the mass
lost in the outburst by the mass transfer rate from the secondary $-\dot{m}_2$ 
\begin{equation}\label{eqn:rec}
t_{\rm quiesc}=\frac{m}{-\dot{m}_2}
\end{equation}
where $m$ is given by \eq{eqn:mass_k}. Hence,
\begin{equation}
t_{\rm quiesc} \propto \frac{K_{\rm max} - K_{\rm min}}{-\dot{m}_2} 
\end{equation}
which gives the ratio of the physical to the simulated quiescence times
as,
\begin{equation}
\frac{t_{\rm quiesc}^{\rm real}}{t_{\rm quiesc}^{\rm sph}} = \frac{K_{\rm max}^{\rm
  real} - K_{\rm min}^{\rm real}}{K_{\rm max}^{\rm sph} - K_{\rm min}^{\rm
  sph}}\left(\frac{-\dot{m}_2^{\rm sph}}{-\dot{m}_2^{\rm real}}\right)
\end{equation}
The recurrence time of the system is simply obtained by $t_{\rm rec}=t_{\rm
  out}^{\rm  real}+t_{\rm quiesc}^{\rm real}$. As will be seen later on, in
the case of \1915, $t_{\rm quiesc}\gg t_{\rm out}$ so that $t_{\rm rec}\sim
t_{\rm quiesc}$.
When interpreting the results based on these scalings, consideration
needs to be given to the assumptions made deriving them. The main
assumption is that $R_{\rm out}$ is independent of the value of
$\Sigma_{\rm max}$. It is 
not exactly the case from one simulation to another, but stays true
within a few percent.

Our canonical simulation has the following parameters $ K_{\rm max}= 55$, 
$K_{\rm min} = 4.79$, $\alpha_h^{\rm sph} = 1$, $\alpha_c^{\rm sph}
= 0.1$, $\eta = 0.1$, $\epsilon = 7\times 10^{-4}$ and $\lambda = 1.0$. 
The values of $\alpha$ are typically a factor of $\sim 10$ larger than the generally 
assumed values $\alpha_c^{\rm real}\approx 0.01$ and $\alpha_h^{\rm
  real}\approx 0.1$.

\section{Results}\label{sec:results}
Outputs from the simulations are presented hereafter. The typical evolution 
of the accretion is described in \S\ref{subsec:typ_beh}. We then explore the 
parameter space of our simulations and report the effects on the disc outburst
and quiescence durations in \S\ref{subsec:alphastuff}, \ref{subsec:sigstuff}
and \ref{subsec:effstuff}. Parameters and outburst/recurrence times of
all the simulations presented below are compiled in table~\ref{table:parameters}.
 \begin{table*}
\begin{center}
\begin{tabular}{cccccccc|cccccccccc}\hline
Simulation & $K_{\rm max}$ & $K_{\rm min}$ & $\alpha_h$ & $\alpha_c$ &
$\eta$ & $\epsilon$ & $\lambda$ &$\dot{M}_1$ & $\dot{M}_{\rm wind} $ &Out. &
Quiesc. & Scaled out. & Scaled quiesc. & duty cycle\\ 
Name & & & & & &($10^{-4}$) & & \multicolumn{2}{c} { ($10^{-5} \msun {\rm
~yr^{-1}}$)} &(days)& (days)& (yr)& (yr) & ($10^{-3}$)\\
\hline
base&		55&	      4.785&	1.0&	0.1&	0.1&	7&	1    &1.4& 0.05&   967 & 868 & 23.8 & 10722 & 2.22\\
									          
irr 1&		55&	      4.785&	1.0&	0.1&	0.1&	3&	1    &1.4 & 0.04&    974 & 873 & 24.0 & 10778 & 2.23\\
irr 2&		55&	      4.785&	1.0&	0.1&	0.1&	5&	1    &1.4 & 0.05&    973 & 877 & 24.0 & 10829 & 2.21\\ 
irr 3&		55&	      4.785&	1.0&	0.1&	0.1&	9&	1    &1.4& 0.05&    978 & 864 & 24.1 & 10673 & 2.26\\
irr 4&		55&	      4.785&	1.0&	0.1&	0.1&	12&	1    &1.3 & 0.07&    1044 & 850 & 26.2 & 10629 & 2.46\\
									          
sig max 1&	40&	      4.785&	1.0&	0.1&	0.1&	7&	1    &1.2 & 0.01&    924 & 582 & 22.8 & 7185 & 3.17 \\
sig max 2&	47.5&	      4.785&	1.0&	0.1&	0.1&	7&	1    &1.4& 0.03&    904 & 769 & 22.3 & 9499 & 2.35\\
sig max 3&	62.7&	      4.785&	1.0&	0.1&	0.1&	7&	1    &1.4 & 0.07&    1016 & 948 & 25.0 & 11715 & 2.14\\
sig max 4&	70&	      4.785&	1.0&	0.1&	0.1&	7&	1    &0.8--1.8&0.04--0.2& 690--1040 & 650--990 & 17--26 & 8029--12229 & 1.39--3.23\\
									          
sig min 1&	55&	      3&	1.0&	0.1&	0.1&	7&	1    &1.3 & 0.02&    1125 & 872 & 27.7 & 10772 & 2.57\\
sig min 2&	55&	      7&	1.0&	0.1&	0.1&	7&	1    &1.5 & 0.08&    839 & 846 & 20.7 & 10454 & 1.98\\
sig min 3&	55&	      10&	1.0&	0.1&	0.1&	7&	1    &1.6 & 0.12&    741 & 800 & 18.3 & 9882 & 1.85\\
sig min 4&	55&	      15&	1.0&	0.1&	0.1&	7&	1    &1.7 & 0.25&    622 & 713 & 15.3 & 8801 & 1.74\\

alpha h 1&	55&	      4.785&	0.5&	0.1&	0.1&	7&	1    &0.9 & 0.04&    1625 & 343 & 40.04 & 4241 & 9.44\\
alpha h 2&	55&	      4.785&	0.8&	0.1&	0.1&	7&	1    &1.3 & 0.06&    1104 & 780 & 27.2 & 9635 & 2.82\\
alpha h 3&	55&	      4.785&	1.3&	0.1&	0.1&	7&	1    &1.6 & 0.16&    733 & 825 & 18.1 & 10191 & 1.77\\
alpha h 4&	55&	      4.785&	1.5&	0.1&	0.1&	7&	1    &1.7 & 0.18&    623 & 717 & 15.3 & 8853 & 1.73\\
									          
alpha c 1&	55&	      4.785&	1.0&	0.05&	0.1&	7&	1    &1.4 & 0.07&    974 & 934 & 24.0 & 21795 & 1.17\\
alpha c 2&	55&	      4.785&	1.0&	0.08&	0.1&	7&	1    &1.4 & 0.05&    971 & 898 & 23.9 & 13716 & 1.74\\
alpha c 3&	55&	      4.785&	1.0&	0.13&	0.1&	7&	1    &1.4 & 0.05&    1022 & 830 & 25.2 & 5565 & 4.53\\
alpha c 4&	55&	      4.785&	1.0&	0.2&	0.1&	7&	1
&**&    ** & ** & ** & ** & ** &**\\
									          
wind 1	&	55&	      4.785&	1.0&	0.1&	0.1&	7&	0.6
& 0.7--1.4 & 0.24--0.90 &605--885 & 625--985& 15--22 & 7720--12167& 1.23--2.85\\
wind 2	&	55&	      4.785&	1.0&	0.1&	0.1&	7&	0.8   &1.3 & 0.21 &   970 & 908 & 23.9 & 11211 & 2.13\\
wind 3	&	55&	      4.785&	1.0&	0.1&	0.1&	7&	1.2  &1.4 & 0.01&   975 & 857 & 24.0 & 10582 & 2.27\\
wind 4	&	55&	      4.785&	1.0&	0.1&	0.1&	7&	1.4
&1.4 & 0.00&   976 & 865 & 24.1 & 10685 & 2.25\\
\\
high irr 1 & 55 & 4.7853 & 1.0 & 0.1 & 0.1 & 28  & 1  & 0.5 & 0.01 & 21500   & ** & 86 & **  &  **\\
high irr 2 & 55 & 4.7853 & 1.0 & 0.1 & 0.1 & 43  & 1  &  0.4 & 0.01 & 22400  & ** & 115 & **  &  **\\
high irr 3 & 55 & 4.7853 & 1.0 & 0.1 & 0.1 & 51  & 1  & 0.4 & 0.02 & 22900 & ** & 130 & **  &  **\\
\hline

\end{tabular}
\end{center}
\caption{Table detailing each setup and results. The parameters used in all
  the simulations are given along with the raw and scaled outburst/quiescence times. 
The parameters where chosen
so that $\dot{M}_1 > - \dot{M}_2$ and the scaling method (see \S\ref{subsec:scalings}) was applicable (for discussion see
\S\ref{subsec:alphastuff}). The setup in {\tt alpha c 4} never allowed the
steady-state to be reached. The last three simulations {\tt high irr [1-3]}
explore large values of $\epsilon $ for direct comparison to \citet{Truss06}
in \S\ref{sec:comp}, but could only be followed during one outburst, hence the lack of
information on the quiescence time (see also discussion in
\S\ref{subsec:effstuff}).
\label{table:parameters}}
\end{table*}
\subsection{Typical behaviour}\label{subsec:typ_beh}
\begin{figure}
  \includegraphics[width=0.52\textwidth]{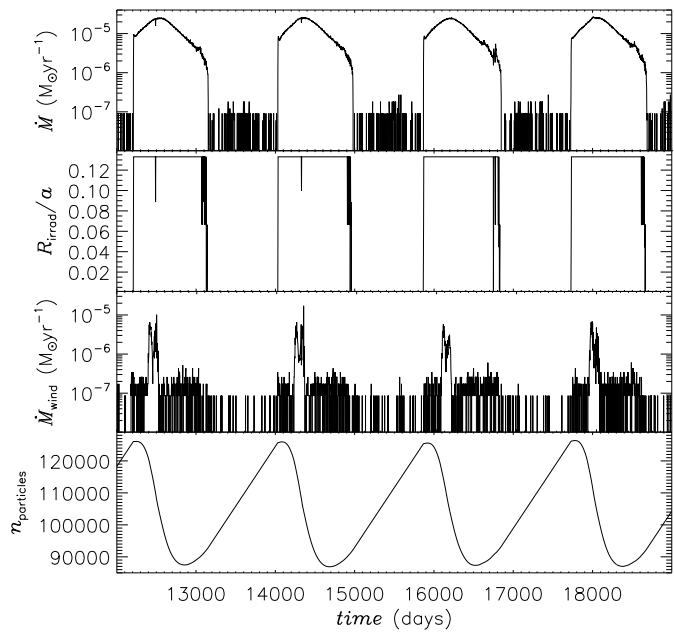}  
  \caption{ From top to bottom: time evolution of the central accretion rate, irradiation radius, wind loss and 
number of  particles in the simulation.\label{fig:dmdt} }
 \end{figure}

As stated in $\S$\ref{subsec:par_inj_rej} particles are
injected from the L1 point and proceed to build up in a disc around
the accreting black hole. We present here a canonical simulation of 
GRS 1915's disc which was run using the parameters recorded in 
the field {\tt  base} of table~\ref{table:parameters}. 
The particle mass and $K_{\rm max}$ are chosen to ensure approximately $10^5$
particles in the quiescent disc. We wait until the system reaches
steady-state over several outbursts\footnote{In this context, {\em steady-state} must be understood as 
the state reached by the disc when the number of particles oscillates around
a constant average value when going through outburst and quiescence phases
(see Fig.~\ref{fig:dmdt}, lower panel).}
to begin our analysis. It is important to emphasise that we do not 
assume any given density profile for the disc but let it build up and find its own 
steady-state. This is in contrast to the analytic treatment of \citet{Truss06}
where the surface density profile of the disc was prescribed by the authors.

The steady-state of the disc is illustrated in Fig.~\ref{fig:dmdt} where
the central accretion rate (top-panel), irradiation radius (second panel),
wind loss (third panel) and number of particles in the simulations (bottom panel)
are plotted against time. 
In the time interval depicted in here, the disc goes through four outbursts
during which the central accretion rate increases by at least three orders
of magnitude up to $\sim 10^{-5}\msun$~yr$^{-1}$. The luminosity of \1915
makes it slightly super-Eddington, and the inferred typical 
accretion rate onto the black hole is $\sim 10^{-7}\msun$~yr$^{-1}$. 
The discrepancy between these two values can be explained by the fact that we
do not resolve the inner disc. The mass transfer of $10^{-5}\msun$~yr$^{-1}$ in
our simulations represents the transfer rate through the disc at $r_{\rm
  in}=0.04a \gg R_{\rm schw}$. 
The amount of mass crossing our inner
boundary is not the amount of mass that will eventually reach the black
hole and we expect most of it to be blown away in a wind. 
In fact, there are good reasons to expect that the mass transfer rate
through such a large disc as is present in \1915 must be much larger than
$\sim 10^{-7}\msun$~yr$^{-1}$
and we discuss that 
point further in \S\ref{subsec:discuss_mdot}.
The accretion in \1915 proceeds at super-Eddington rates and this explains the
plateau observed in $R_{\rm irr}$ as the latter is capped at the Eddington
radius (see \S\ref{subsec:irradiation}). This is a general
feature of all our simulations.

Caution is needed when reading outburst and quiescence durations
from this figure as the plotted time is a direct output from the code: outburst
and recurrence times have to be scaled according to
\S\ref{subsec:scalings}. 
Doing so, one gets $t_{\rm out}=24$~yrs and $t_{\rm quiesc}=10722$~yrs for this 
particular simulation.

\begin{figure*}
  \centering
  \includegraphics[width=0.33\textwidth]{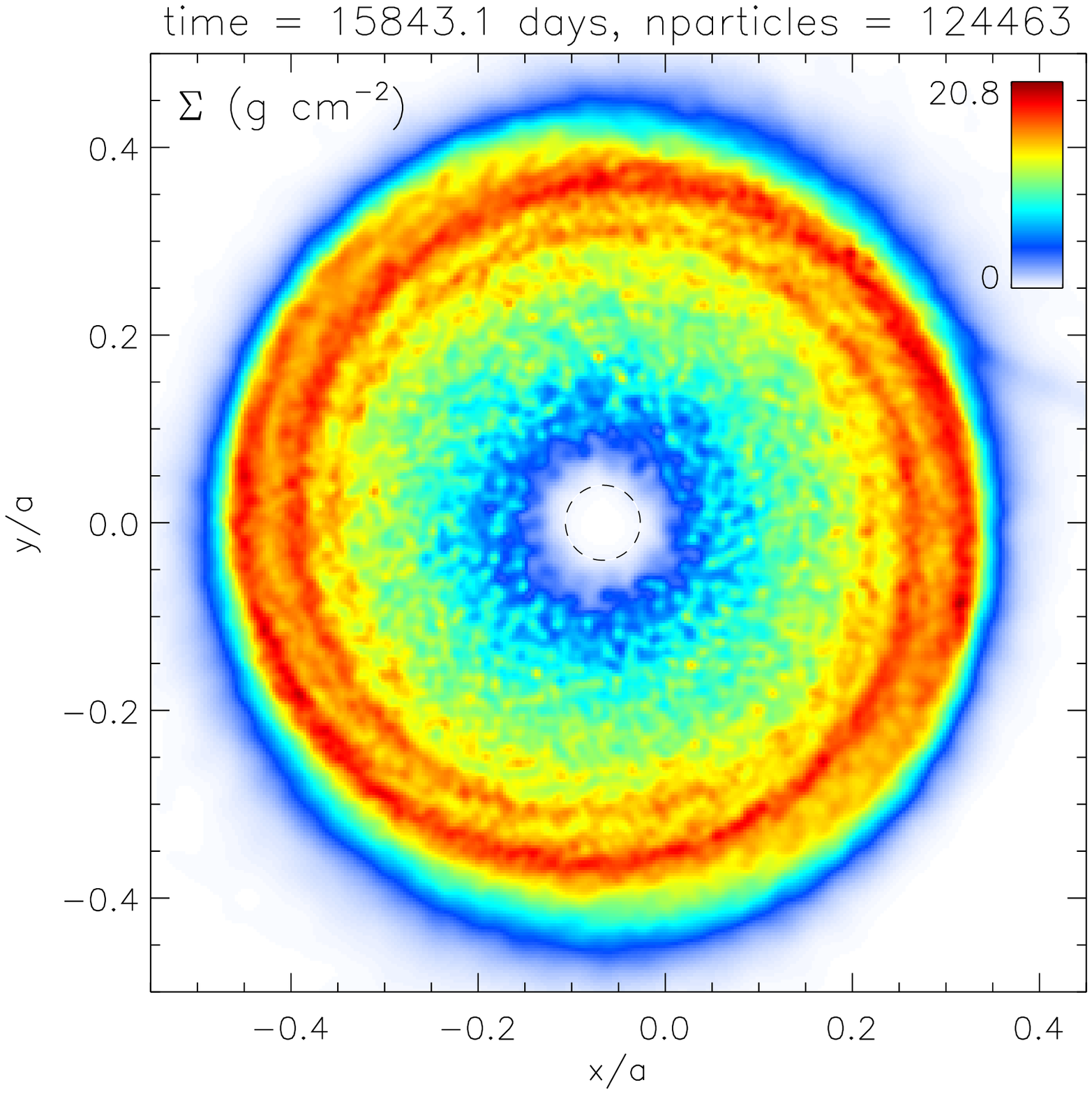}  
  \includegraphics[width=0.33\textwidth]{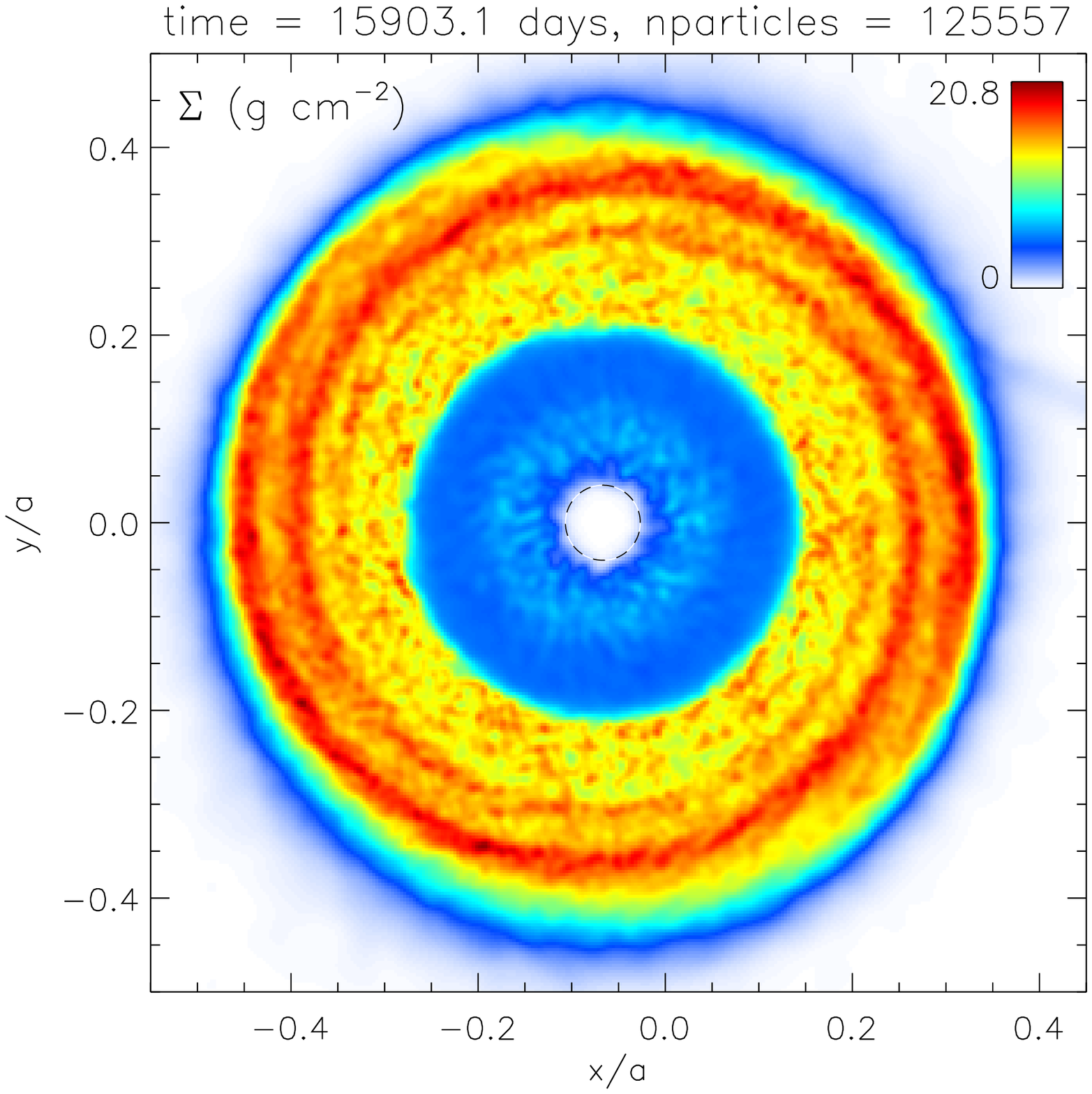} 
 \includegraphics[width=0.33\textwidth]{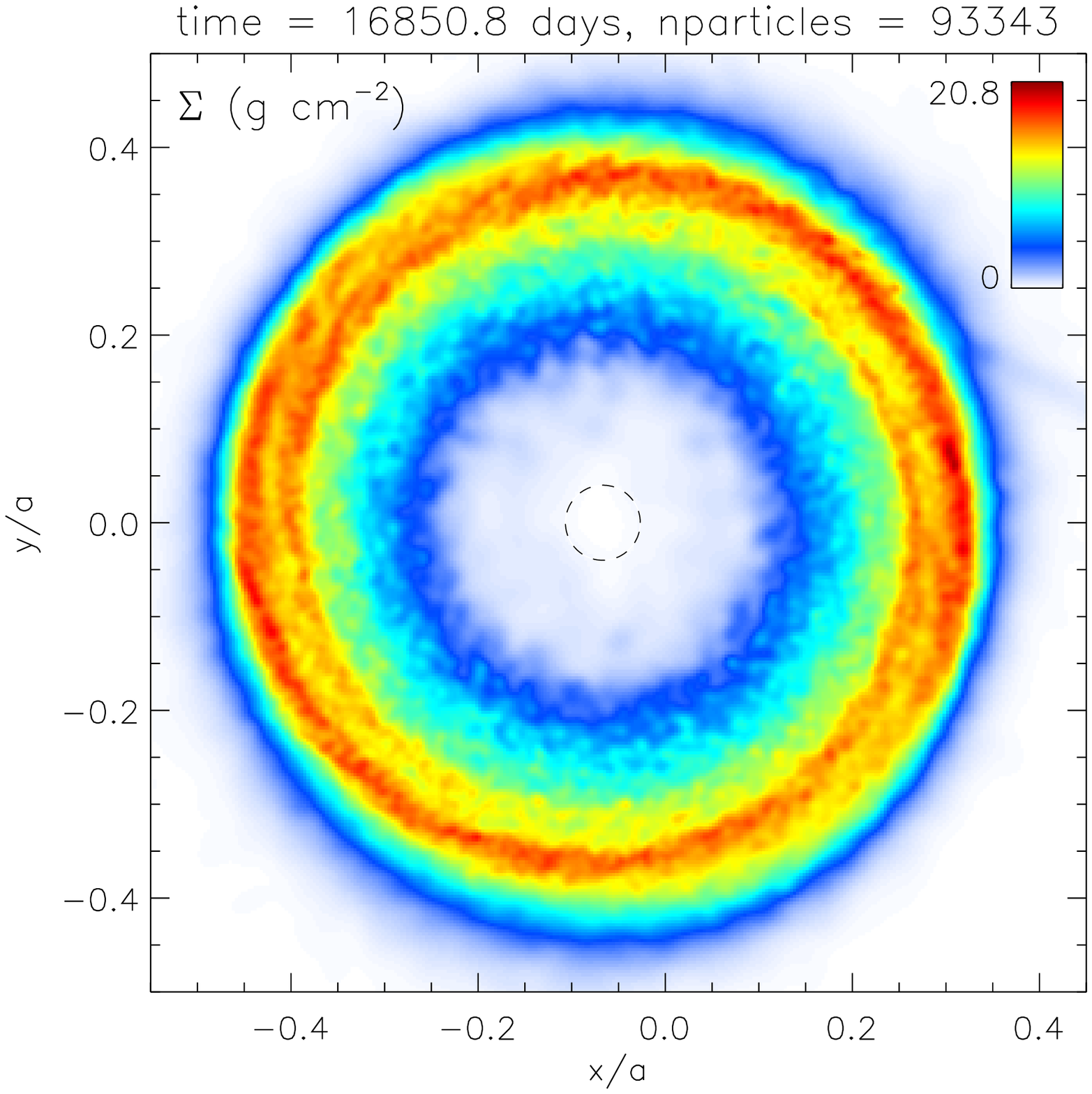} 
  \caption{Snapshots of the surface density of the disc before (left), during
    (middle) and after (right) an outburst. The dashed circle represents the
    inner radius of the simulation.\label{fig:pretty_pics}}
  \label{fig:out_evol}
 \includegraphics[width=0.33\textwidth]{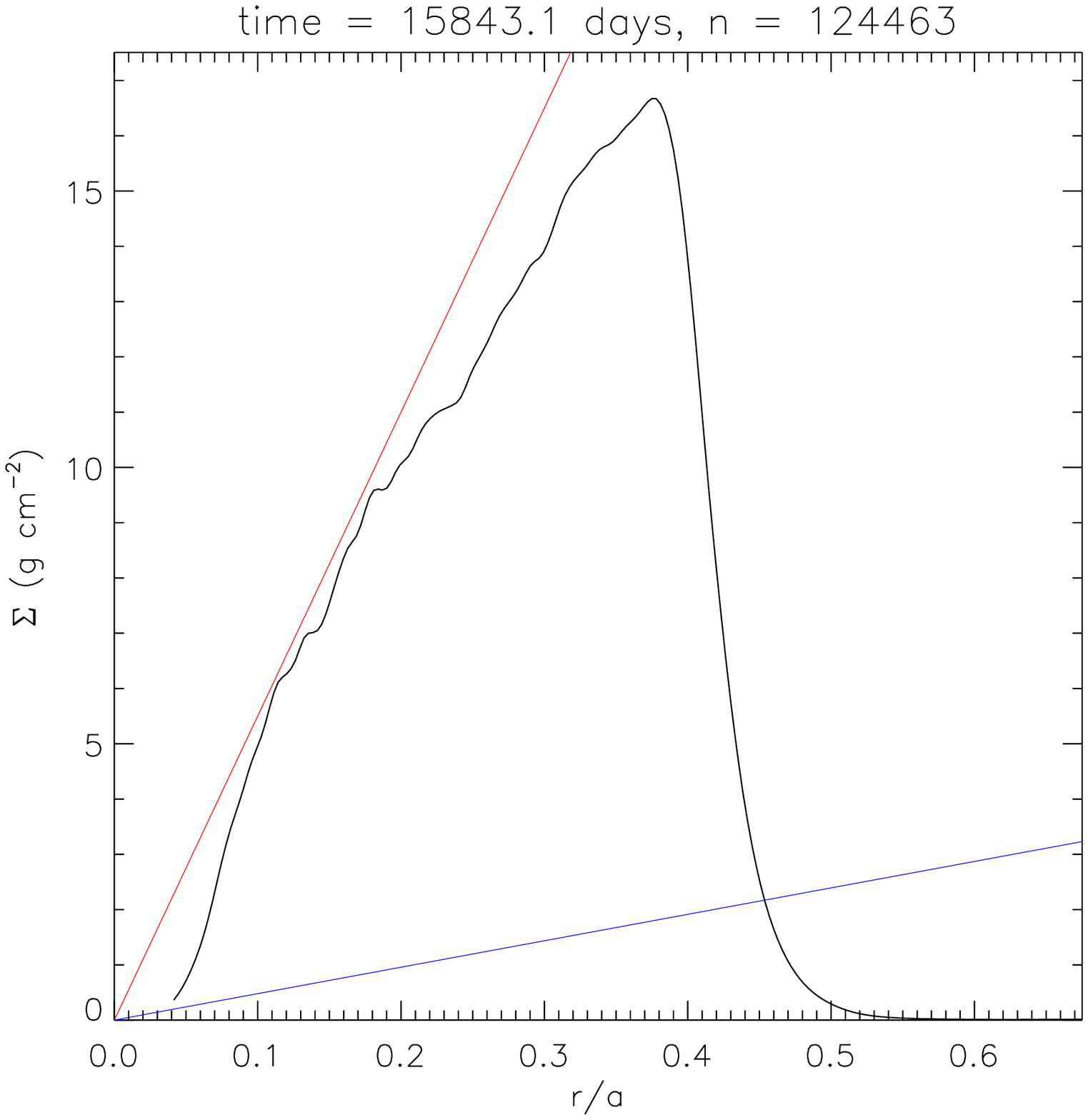}               
\includegraphics[width=0.33\textwidth]{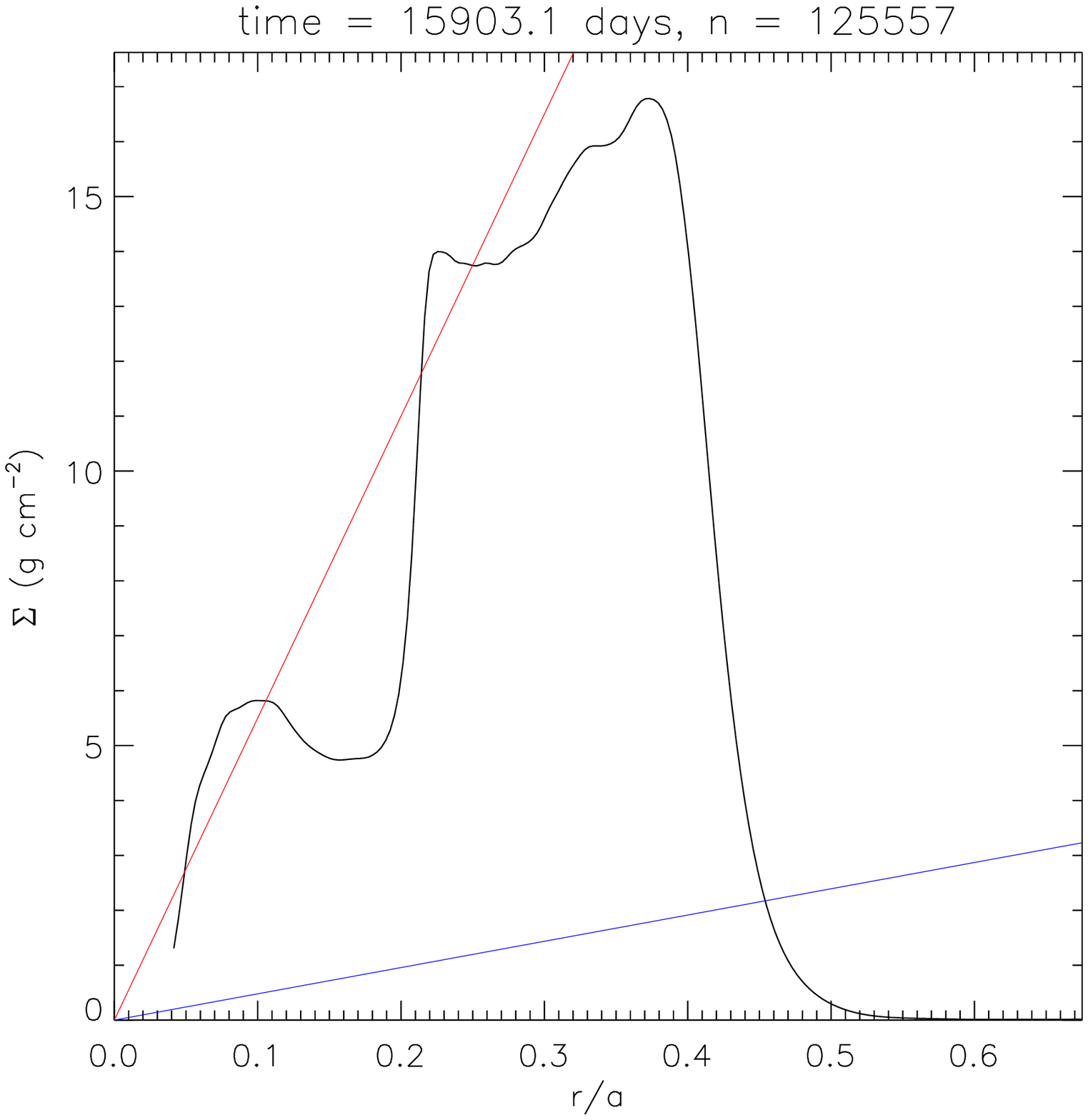}
 \includegraphics[width=0.33\textwidth]{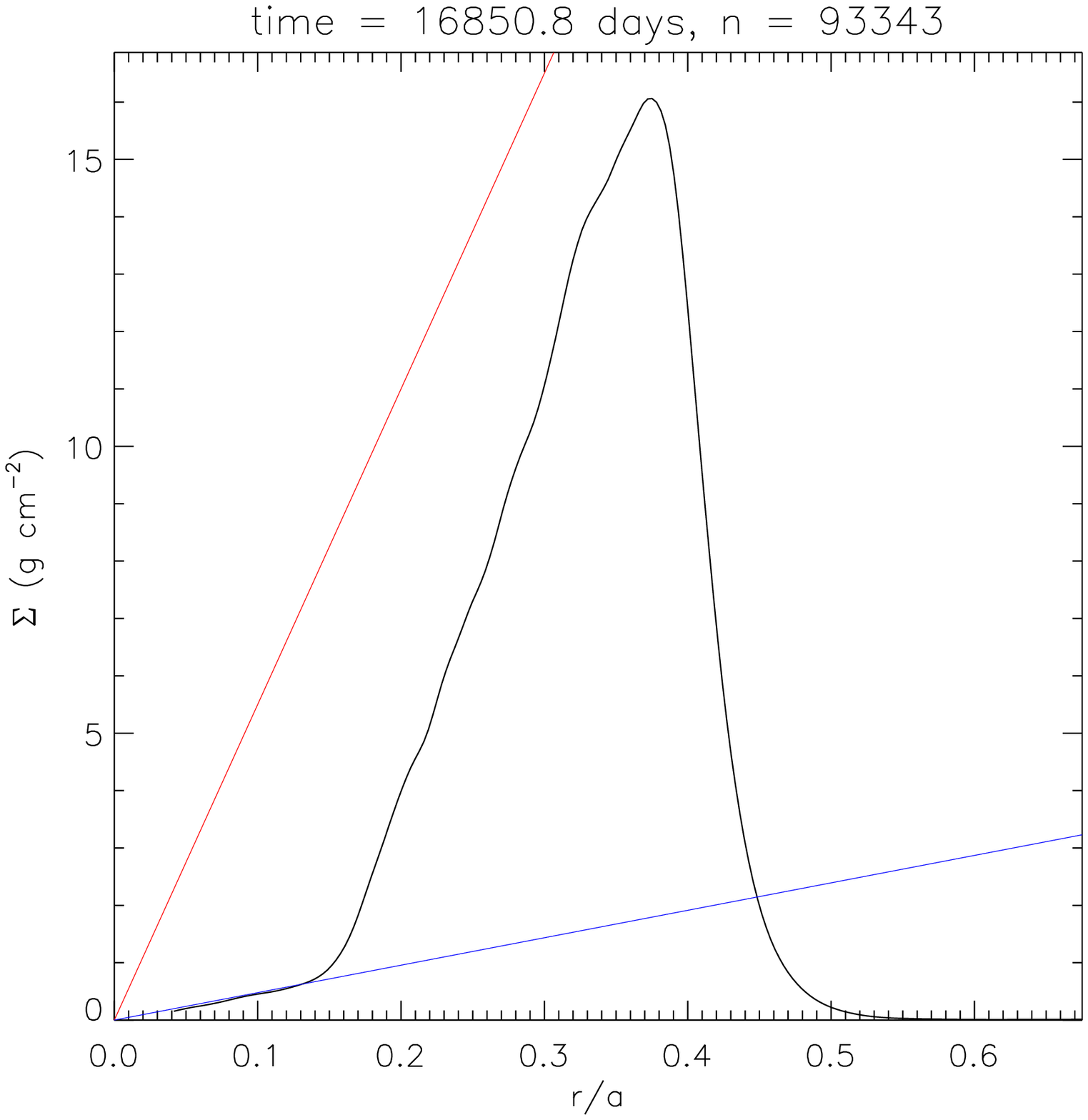} 
\caption{Evolution of the azimuthally averaged surface density before (left), during
    (middle) and after (right) the outburst of fig.~\ref{fig:pretty_pics}. Red and blue lines
 represent the critical surface densities $\Sigma _{\rm max}$ and $\Sigma _{\rm min}$. \label{fig:disc_evol}}
\end{figure*} 

Figures~\ref{fig:pretty_pics} and \ref{fig:disc_evol} follow the evolution of the surface
density in the disc for the {\tt base} simulation through an
outburst. In fig.~\ref{fig:disc_evol}, the surface density at a given radius has been averaged
over 2$\pi$.  In each case, the three snapshots (a), (b) and (c) are taken just before, during,
and after the outburst respectively. 

\begin{figure}
  \includegraphics[width=0.5\textwidth]{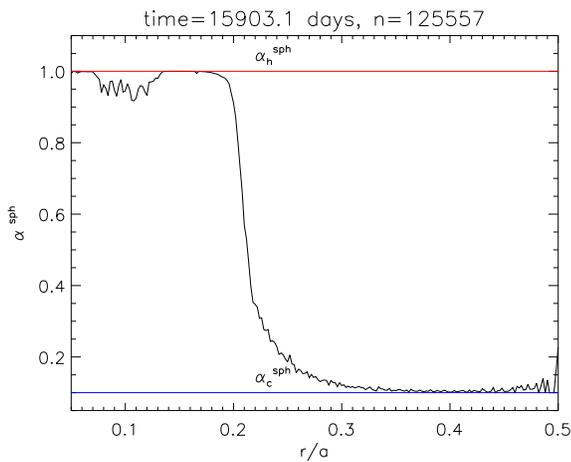}
  \caption{Radial profile of $\alpha$ at the beginning of the outburst phase. The
    inner disc is switching to the hot state while the outer regions remain
    in at low viscosity.}
  \label{fig:alpha}
\end{figure}

In fig.~\ref{fig:disc_evol}a, the surface density in the inner disc closely
follows the $\Sigma_{\rm max}(r)$ prescription. 
Exceeding this limit at $r \sim 0.2a$ triggers the switch to the hot viscous
state and marks the beginning of the outburst
phase (see \S\ref{subsec:inst_num}). The switch in viscosity is
illustrated in fig.~\ref{fig:alpha} showing the radial variation of the SPH
$\alpha$ parameter at the start of the outburst.
The particles in the annulus that enter the hot state spread to increase the surface density in
neighbouring annuli and have them switch to the high viscosity state as
well (fig.~\ref{fig:disc_evol}b). Because of these matter waves propagating 
inward and outward, a significant portion
of the disc enters the hot state, to be eventually accreted onto the 
black hole (but for the part blown away in the wind).

Typically 20-30\% of the particles (i.e. of the mass of the disc) are accreted
during an outburst (Fig.~\ref{fig:dmdt}, lower panel). 
Once most of the material has been accreted, 
the surface density in the inner disc decreases to eventually become
smaller than $\Sigma_{\rm min}$, as seen in fig.~\ref{fig:disc_evol}c. 
This is the trigger for the disc to switch back to the cold low-viscosity
state and ends the outburst phase.
During quiescence the disc regains the mass it lost in outburst and the cycle repeats. 

Figure~\ref{fig:out_evol} is another representation of the above, where 
one clearly sees the inner disc being emptied throughout the outburst.
The 2D plots have also the benefit to show the existence non axi-symmetric features, 
e.g. large spiral arms developing in the outer disc.

\subsection{Varying the viscosity}\label{subsec:alphastuff}
\begin{figure}
\begin{center}
\includegraphics[width=9cm]{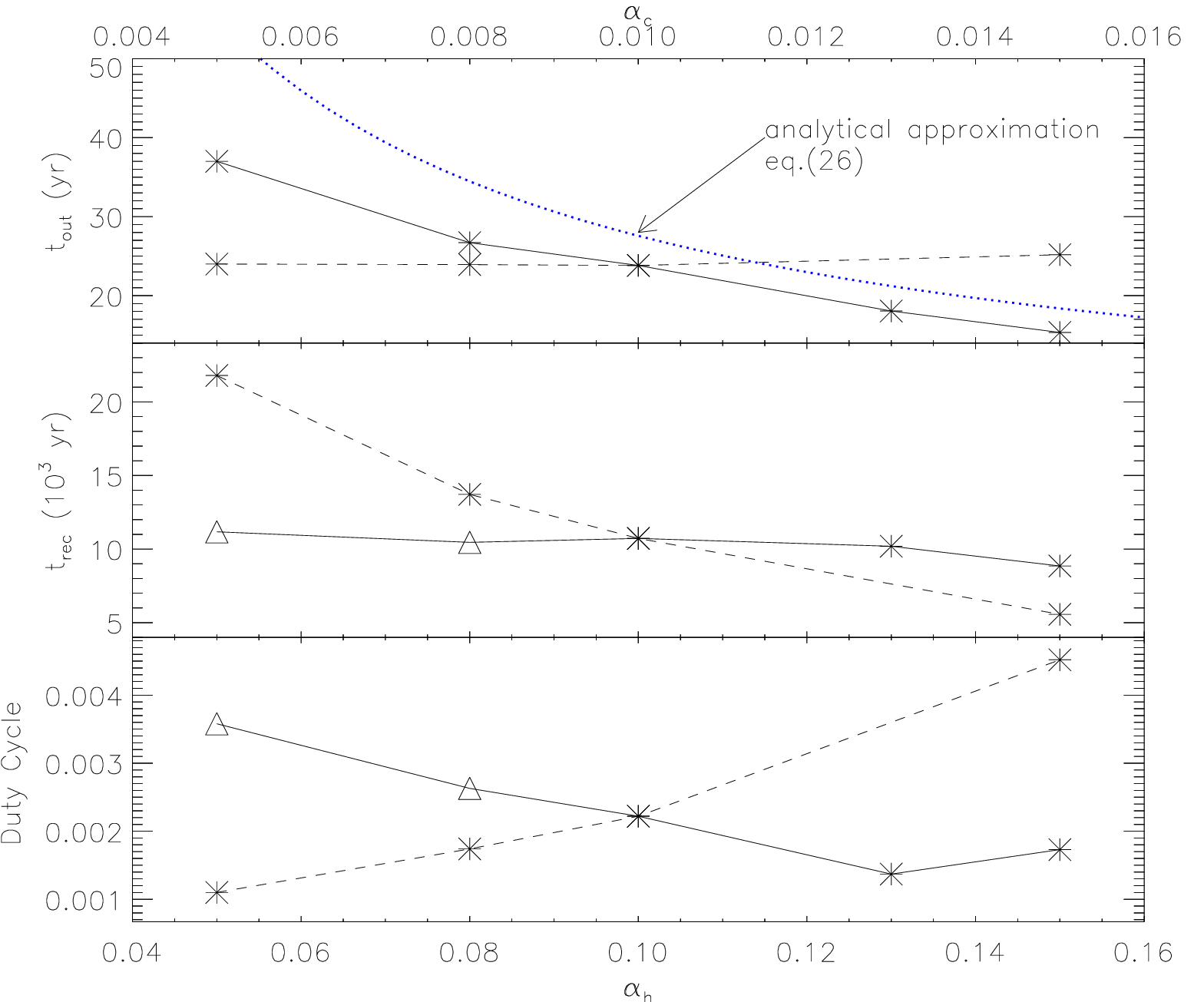}
\caption{$t_{\rm out}$, $ t_{\rm rec}=t_{\rm out}+t_{\rm quiesc}$ and duty cycle shown as a function of
  $\alpha_h$ (solid line) and $\alpha_c$ (dashed line). The two triangles
  represent simulations that have not been scaled by the method described in
  \S\ref{subsec:scalings}. See text for details. }
\label{fig:alpha_h_c}
\end{center}
\end{figure}

The $\alpha$-parametrisation of the viscosity in accretion discs has been 
widely used since \citet{shakura73}. However, to date some confusion still 
exists around the actual value(s) that $\alpha$ should take to describe the
angular momentum transport in accretion discs \citep{2007MNRAS.376.1740K}. In this section,
we allow for $\alpha_c$ and $\alpha_h$ to depart from their canonical values and
report their effect on the disc's limit cycle. The following results have been 
obtained from the simulations {\tt alpha h [1-4]} and {\tt alpha c [1-4]} which full set of parameters 
can be found in tab.~\ref{table:parameters}. For numerical reasons, we could
not change $K_{\rm max/min}^{\rm sph}$ with $\alpha_{c/h}$ as
eqs.~(\ref{eqn:sigmamax}) and (\ref{eqn:sigmamin}) require. Indeed, changing
$K_{\rm max/min}^{\rm sph}$ on a large range of values results in the loss of
the transient behaviour we are studying here. Nevertheless, the scaling of 
the outburst and recurrence times accounts for this technical detail.
Figure \ref{fig:alpha_h_c} shows the effect of varying $\alpha_h$ (solid line)
and $\alpha_c$ (dashed line) on $t_{\rm out}$, $t_{\rm quiesc}$ and duty cycle. Note that the values of
$\alpha$ displayed are the physical values and not the SPH ones ($\alpha_{
  h/c}=\alpha_{ h/c}^{\rm sph}/10$). The timescales are scaled as described in
\S\ref{subsec:scalings} if not stated otherwise. 

The effects of changing $\alpha_c$ are straightforward: little effect 
if any on the outburst time (fig.~\ref{fig:alpha_h_c}, top panel, dashed line)
as one would expect and an increasing recurrence time with decreasing
$\alpha_c$ (middle panel). Reducing $\alpha_c$ implies an increase of
$\Sigma_{\rm max}$ according to \eq{eqn:sigmamax}. Assuming that the 
outer radius of the outburst and the density profile of the disc are only
slightly affected by this change\footnote{This is verified in
  \S\ref{subsec:sigstuff}. 
The trigger point of the outburst changes with the critical density profile but the radius
to which the hot state propagates is roughly unchanged.} then
more mass should be lost in the outburst for a smaller $\alpha_c$. The
transfer rate from the secondary being fixed, this translates into longer
quiescence times.

The outburst time decreases with increasing $\alpha_h$ as accretion proceeds much faster
for larger values of the viscosity (fig.~\ref{fig:alpha_h_c}, top panel, solid
line). As a result, the mass rate at the inner boundary increases making the
flow even more super-Eddington. This however, does not translate into more
mass being lost in the outburst. The number of particles (not shown here) 
actually lost during an outburst decreases as $\alpha_h$ increases. With high
values of $\alpha_h$, the inner disc empties quickly and quenches the 
outburst before the outward matter wave could turn more of the disc into the 
hot state.
The outburst time is roughly governed by the viscous time calculated at the
outermost part of the outburst $R_{\rm out}$ and reads \cite{Pringle81},
\begin{eqnarray}\label{eqn:out_time}
t_{\rm out} &\sim& t_{\rm visc} =  \frac{1}{\alpha_h \Omega_K (H/R)^2}
\\ \nonumber
&\approx& 170 \left(\frac{\alpha_h}{0.1}\right)^{-1}
\left(\frac{(H/R)_{\rm out}}{0.03}\right)^{-2} \left(\frac{R_{\rm
    out}}{a}\right)^{3/2} {\rm yrs}\;.
\end{eqnarray}
where $\Omega_K = (GM_1/R^3)^{1/2}$ has been used. The simulations give at most 
$H/R \sim 0.03$ and $R_{\rm out}\sim  0.3a$, which gives 
\beq
t_{\rm out} \sim 2.8/\alpha_h\;{\rm yrs}.
\eeq 
This simple analytical estimate is a rough upper limit for the outburst time and 
is plotted as the dotted blue line in fig.~\ref{fig:alpha_h_c} (top panel)
where it can be seen to reasonably reproduce the
behaviour of the numerical results.

The recurrence time also marginally decreases with increasing $\alpha_h$. This is
understood when remembering that the quiescence phase of the disc is roughly
the time needed to replenish the mass lost in the outburst at a rate $-\dot M_2$. As mentioned 
earlier, more mass is lost in the outburst for smaller values of $\alpha_h$
which explains the trend reported in fig.~\ref{fig:alpha_h_c} (middle panel,
solid line). Note that the two triangle points have not been scaled using the
method reported in \S\ref{subsec:scalings}. Indeed, the latter is only valid
if $\dot M_1 \gg -\dot M_2$. This would always be the case if working with
the real disc mass, but because of the numerical setup (small $K_{\rm max}$ and
$K_{\rm min}$), we only get $\dot M_1
\gtrsim -\dot M_2$ for the lowest values of $\alpha_h$. This means that, for
these two simulations, the disc significantly replenishes while still in
outburst. This indicates the disc is close to persistent outburst.
However, it is a limitation of the simulations and limits the range over which parameters are allowed to vary.
The duration measured from  the evolution plots (like that
of fig.~\ref{fig:dmdt}) largely underestimates the actual quiescence times.
To overcome that issue, the quiescence times of these two points have been
explicitly calculated from  $t_{\rm quiesc}=M_{\rm out}/-\dot M_2$, 
where $M_{\rm out}$ is the total mass lost in the outburst and not from the
raw value of $t_{\rm quiesc}$ output from the code.

The lower panel in fig.~\ref{fig:alpha_h_c} shows the variation of the duty 
cycle. Both $\alpha_c$ and $\alpha_h$ affect it in opposite ways, similar 
in amplitude. Nonetheless, it spans on a very narrow range with an average value
from 0.1\% to 0.45\%.

\subsection{Changing the critical density profiles \label{subsec:sigstuff}}
\begin{figure}
\includegraphics[width=9cm]{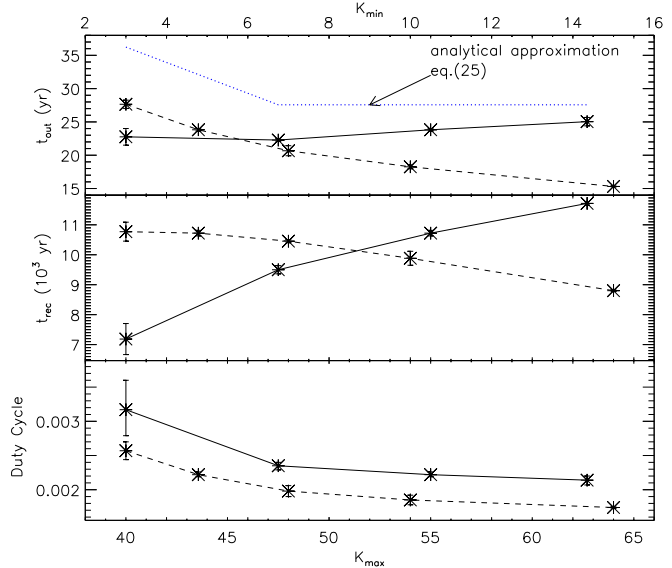}
\caption{$t_{\rm out}$, $\rm t_{\rm rec}$ and duty cycle as a function of
  $K _{\rm max}$ (solid line) and $K_{\rm min}$ (dashed line).}
\label{fig:sig_max_min}
\end{figure}

The critical density profiles we use are given by \eq{eqn:gradmax} and
(\ref{eqn:gradmin}).
which are fitted on the vertical disc structure obtained by
\citet{cannizzo88}. Different fits can be obtained depending on the disc model
(irradiated or not) used and other prescriptions can be found in the literature 
\citep{1984ApJS...55..367C,1998MNRAS.298.1048H,dubus01}. Our setup does not
allow for irradiation feedback on the disc structure or changing 
$\Sigma_{\rm max/min}$ accordingly. As also mentioned in the previous section, 
we do not have much freedom to change their normalisations $K_{\rm max/min}$ and
still retain the transient behaviour of the system. 
Within the range of values where the transient behaviour is present, 
fig.~\ref{fig:sig_max_min} displays the effects of 
$K_{\rm  max}^{\rm sph}$ (solid line) and $K_{\rm min}^{\rm sph}$ (dashed line) 
parameters on the outburst and recurrence times and on the duty cycle 
(simulations {\tt sig~max~[1-3]} and {\tt sig~min~[1-4]} in table~\ref{table:parameters}).

Increasing $K_{\rm min}^{\rm sph}$ implies a steeper slope for the critical surface
density $\Sigma_{\rm min}^{\rm sph}$. As the outburst proceeds and the surface density
in the inner disc diminishes, the latter falls below $\Sigma_{\rm min}^{\rm sph}$
quicker for higher values of $K_{\rm min}^{\rm sph}$ (also see
fig.~\ref{fig:disc_evol}c). The outburst time is then smaller for higher
values of $K_{\rm min}^{\rm sph}$ (fig.~\ref{fig:sig_max_min}, top panel, dashed line).
The mass lost in the outburst is decreased in a similar fashion (see \eq{eqn:mass_k}) so that 
the time needed to replenish the disc also decreases with 
increasing $K_{\rm min}^{\rm sph}$ (middle panel, dashed line).

\begin{figure}
\includegraphics[width=4.2cm,angle=90]{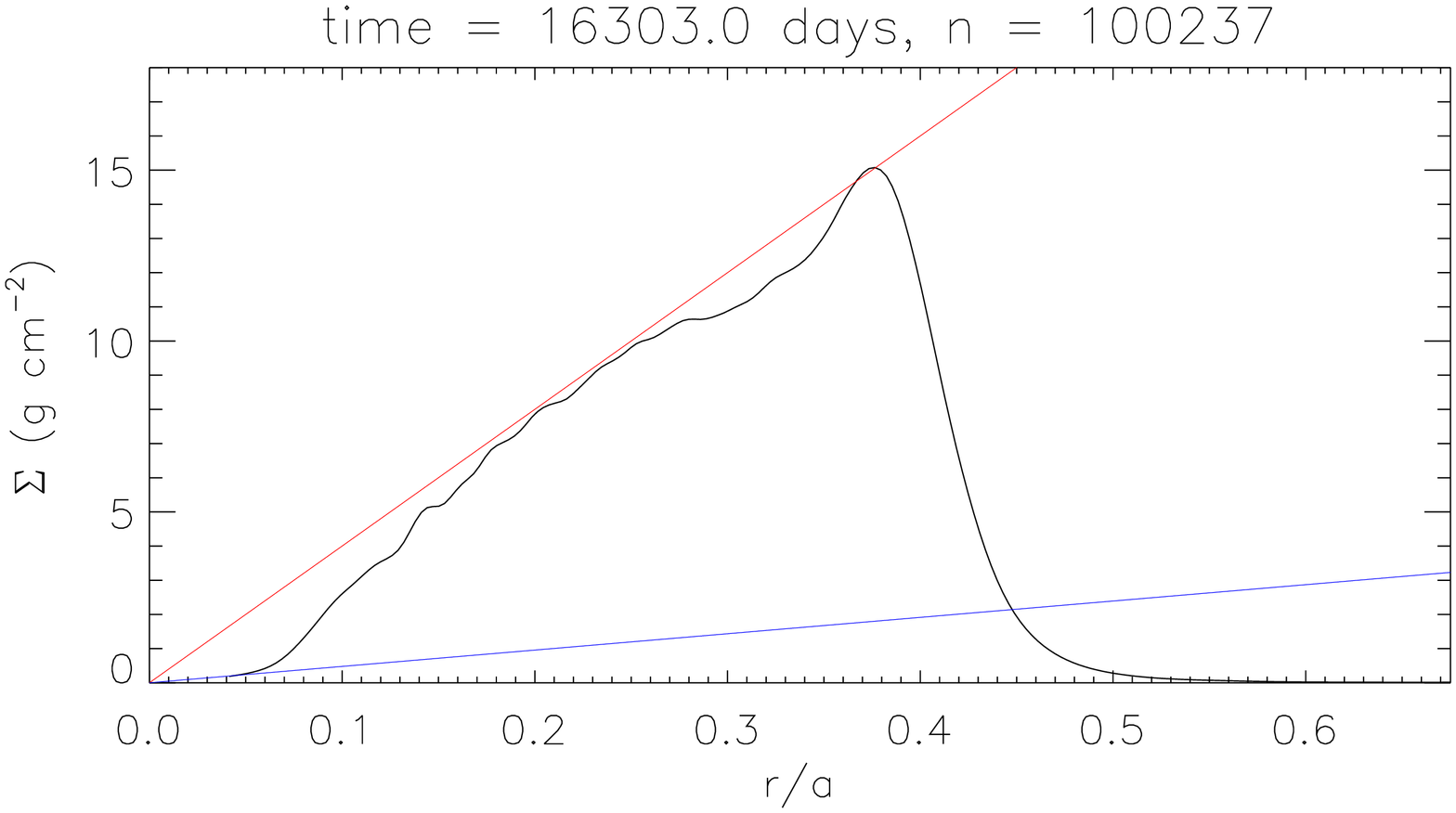}\vspace{+5pt}
\includegraphics[width=4.2cm,angle=90]{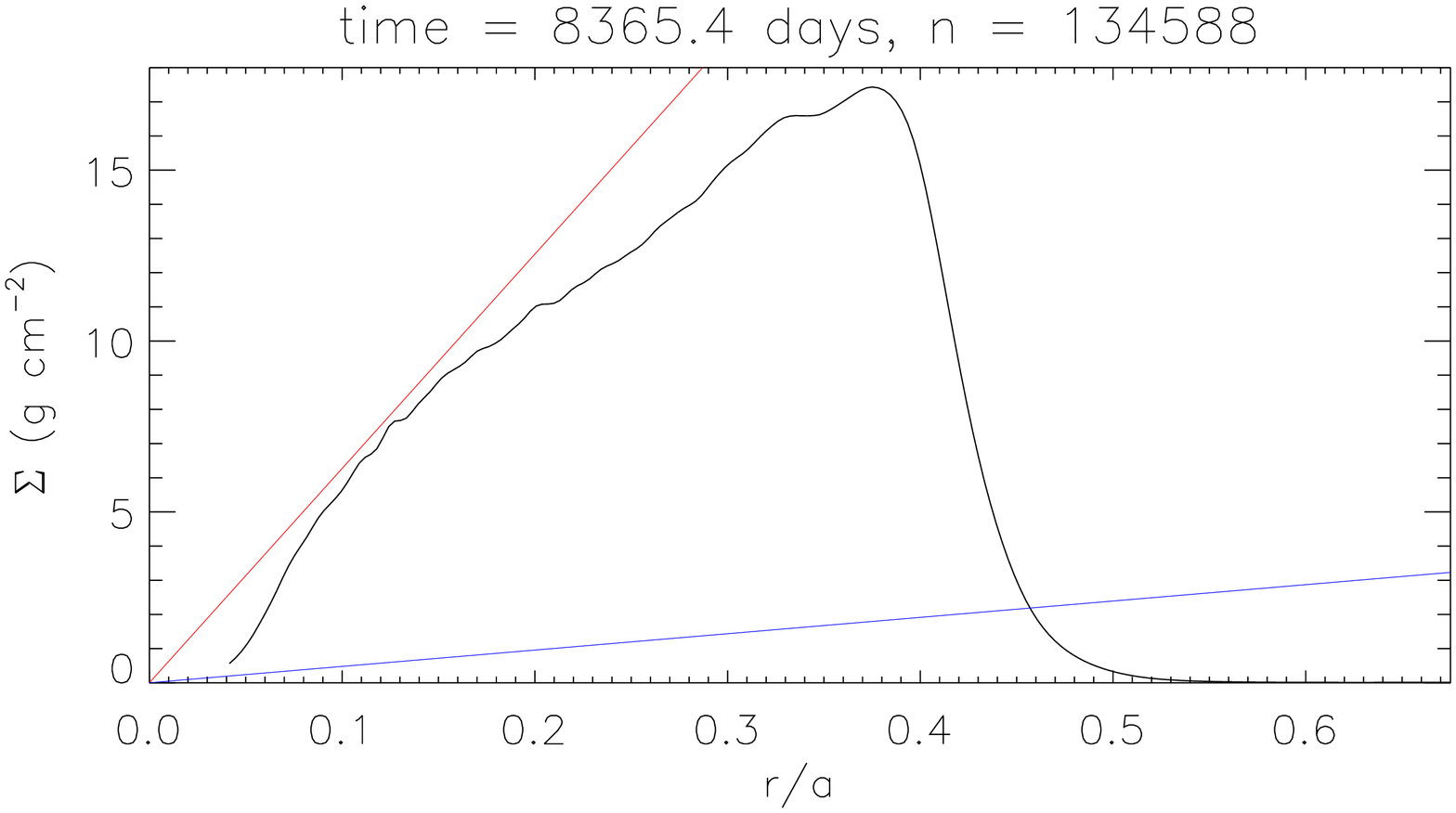}
\caption{Disc density profiles just prior outburst (black solid line). \emph{Top:}
  simulation with $K_{\rm max}^{\rm sph} = 40$~g~cm$^{-2}$. 
\emph{Bottom:} simulation with $K_{\rm max}^{\rm sph} = 60$~g~cm$^{-2}$. 
$K_{\rm min}^{\rm sph}$ is the same in both cases. The triggering radius of
the outburst is small for larger $K_{\rm max}^{\rm sph}$. \label{fig:sig_max_comp}}
\end{figure}

The behaviour of the outburst and recurrence times with varying $K_{\rm
  max}^{\rm sph}$
is less straightforward. First of all, a steeper slope of $\Sigma_{\rm
  max}^{\rm sph}$
means a smaller trigger radius for the outburst. This is shown in 
fig.~\ref{fig:sig_max_comp} where the snapshots are taken just before 
the outburst. In the upper panel, $K_{\rm max}^{\rm sph}$ has a low value:
the outburst is triggered in the outer disc, at $r\sim 0.38a$,  
and will propagate inward (outside-in outburst). For a high value of 
$K_{\rm max}^{\rm sph}$ (lower panel), the surface density crosses its critical value
at $r\sim 0.12a$ and the front will propagate outward, up 
to $R_{\rm out} \sim 0.3a$  
(inside-out outburst, similar to that of fig.~\ref{fig:disc_evol}). 
In an inside-out outburst, the outer radius of the 
outburst does not appear to depend on $K_{\rm max}^{\rm sph}$: this is the
case for the {\tt  base} and {\tt sig~max [2-3]} simulations. In the latter, the hot state 
always propagates up to $R_{\rm out}\sim 0.3a$ so according to \eq{eqn:out_time},
the outburst time should remain unchanged. This is seen in
fig.~\ref{fig:sig_max_min} (top panel, solid line). The only deviation from
the analytical upper estimate is for the lowest value of $K_{\rm max}^{\rm sph}$
where the outburst propagates outside-in and is triggered at $R_{\rm out}\sim
0.38a$. Used in \eq{eqn:out_time}, it gives an outburst duration much larger
than the simulated one. This indicates that this analytical estimate 
breaks down for outside-in outbursts. With increasing $K_{\rm max}^{\rm sph}$,
more mass is lost in the outburst, regardless of its type, and the quiescence
time increases accordingly (middle panel, solid line).

The duty cycle decreases with both $K_{\rm max}^{\rm sph}$ and 
$K_{\rm  min}^{\rm sph}$. Again, it spans on a narrow range around 0.25\%.

\subsection{Effect of the irradiation and wind efficiency \label{subsec:effstuff}}

\begin{figure}
\includegraphics[width=9cm]{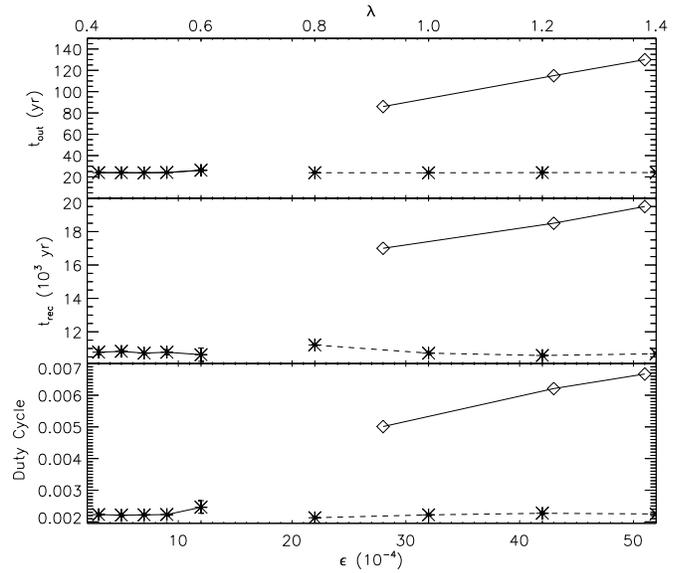}
\caption{$t_{\rm out}$, $\rm t_{\rm rec}$ and duty cycle shown as a function of
  $\epsilon$ (solid line) and $\lambda$ (dashed line). Crosses are for
  simulations where the particle injection is on during the outburst whereas
it is turn off for the diamond simulations. See text for details.}
\label{fig:irr_wind}
\end{figure}

The last two free parameters of our model are the irradiation and wind
efficiency, $\epsilon$ and $\lambda$ respectively. 
The former defines the irradiation radius given by \eq{eqn:irr} and 
the latter the threshold mass rate for the wind to be active\footnote{The
  higher $\lambda$ the larger the threshold, hence the less efficient the
  wind.} as established by \eq{eqn:wind_threshold}. The irradiation efficiency depends on the 
disc structure\footnote{The disc structure itself depends on the irradiation
heating.}  and albedo of the gas (see \citet{Frank02}), two quantities 
our model does not include and we treat $\epsilon$ as a free parameter. 
There is evidence for slow winds being emitted from accretion
discs in LMXBs \citep{2006Natur.441..953M,2009arXiv0901.1982U}. However,
our code does not possess the ingredients for the disc to consistently 
emit a wind, given its thermodynamical properties. Varying $\lambda$ allows
us to artificially give more or less importance to the wind without resolving
its origin.
Figure \ref{fig:irr_wind} illustrates the impact of varying these two
parameters on the temporal evolution of the disc (simulations {\tt irr~[1-4]}
and {\tt wind [1-4]} from table~\ref{table:parameters}). 

The efficiency of the wind has no noticeable effect on
the outburst time (fig.~\ref{fig:irr_wind}, top panel, dashed line). 
Indeed, $R_{\rm out}$ is not changed significantly 
whatever $\lambda$ so that the viscous time in the hot state remains the 
same. The rate at which mass is lost in the wind is roughly equal to the 
central mass accretion rate but these high rates lasts only during a small  
fraction of the outburst duration (fig.~\ref{fig:dmdt}, top and third panel):
the mass lost in the wind is much smaller than that accreted during 
an outburst. Therefore, the recurrence time does not show any significant change,
as long as $\lambda$ is high enough. For the smallest value of $\lambda$
however, the threshold for the wind is so low that a wind is emitted during
quiescence. It then takes longer to replenish the disc, hence the slight
increase of $t_{\rm rec}$ when $\lambda=0.8$. The main effect of the wind
during outburst is to reduce the central accretion rate (not shown here).

Changing the irradiation efficiency leads to much more striking behaviour
(solid lines, crosses and diamonds). The system always presents super-Eddington accretion so that
the irradiation radius is capped at Eddington and simply given by
\eq{eqn:redd}. The effect of the irradiation is twofold: i) it maintains the 
irradiated region in the hot state, even if $\Sigma(r)<\Sigma_{\rm min}$, 
ii) it can switch external regions where $\Sigma(r)<\Sigma_{\rm max}$ into the hot
state.
Both effects should tend to increase the outburst time.
For low values, up to $\epsilon \sim 10^{-3}$, $R_{\rm irr} < R_{\rm out}$ and only the
first effect is at play (top panel, solid line, crosses). That regime only
affects a small amount of mass and the outburst duration plateaus for
$\epsilon <10^{-3}$ and increases only when $\epsilon \sim 10^{-3}$. The
scale of the graph makes it difficult to visualise and we refer the reader
to table~\ref{table:parameters} for the actual values.

To explore the second regime, $\epsilon>10^{-3}$ is required. Because
of the scaled-down disc we simulate, pushing $\epsilon$ to higher values
leads to $M_1 \sim -\dot{M}_2$ and leads the disc to be in a continuous outburst state.
To overcome that difficulty, we turn off the injection of particles during the
outburst phase ($\dot{M}_2=0$) in order to get the outburst to come to an end
(this corresponds to the simulations {\tt high irr [1-3]} in table~\ref{table:parameters}).
This is a valid approach since \1915 has $\dot M_1 \gg -\dot{M}_2$. In
fig.~\ref{fig:irr_wind} these simulations are represented by the diamond symbol.
In that regime, $R_{\rm out}=R_{\rm irr}$ is much larger and a significant
amount of mass enters 
the hot state in the outer disc, resulting in much longer outbursts.
Remembering that $R_{\rm irr}\propto \epsilon^{1/2}$ and $t_{\rm
  out}\sim t_{\rm visc}\propto R_{\rm out}^{3/2}$, one gets $t_{\rm out}\sim
\epsilon^{3/4}$.
This is in reasonable agreement with the behaviour followed by the three diamond
simulations, where $t_{\rm out}\propto \epsilon^{0.7}$.
For the highest values of $\epsilon$ the entire disc is irradiated and 80\% of 
its mass is lost in the outburst.

The recurrence time follows directly from there. The longer the outburst
time, the more mass there is to replenish, hence the longer the 
recurrence time.

\section{Discussion}\label{sec:discussion}

\subsection{ The outburst duration \label{sec:comp}} 

\citet{Truss06} analytically estimated the outburst duration
of \1915. We briefly outline the framework of their calculation before 
comparing to our results. 
In their work, the outburst time was calculated 
following

\begin{equation}
t_{\rm out} = \frac{M_{\rm out}}{\langle\dot{M}_{\rm disc}\rangle}.
\end{equation}
The surface density in the disc was
taken to follow $\Sigma_{\rm max}$ in the inner 10\% of the disc before
flattening off at larger radii. Hence, 
\begin{equation}
M_{\rm out} = \int_{0.1R_{\rm disc}}^{R_{\rm out}}2\pi R \Sigma_{\rm
max}(0.1R_{\rm disc}){\rm dR},
\end{equation}
where $R_{\rm out}$ was simply the irradiation radius
(\eq{eqn:irr}). 
The average mass transfer rate through the disc read
\begin{equation}
\langle\dot{M}_{\rm disc}\rangle = \langle\dot{M}_{1}\rangle +
\langle\dot{M}_{\rm wind}\rangle - \dot{M}_2.
\end{equation}
Remembering that $\dot{M}_2 \ll \langle\dot{M}_{1}\rangle$ and using $\dot
M_1=\dot M_{\rm Edd}$, the inclusion of
$\langle\dot{M}_{\rm wind}\rangle$ as free parameter resulted in a maximum and a minimum length
of an outburst for a given irradiation efficiency\footnote{This cannot be directly
compare to our set up, where we showed that the efficiency of the wind had
little influence on the outburst time (fig.~\ref{fig:irr_wind}, top panel, dashed line). 
It is difficult to make an analogy between our two approaches,
but we can understand this apparent discrepancy in the following way: in our
simulation, the measured $\dot M_1$ in $r_{\rm in}\gg R_{\rm Schw}$ \emph{knows} of
the existence of a wind (i.e. the stronger the wind, the smaller $\dot M_1$)
so that overall $\langle\dot{M}_{\rm wind}\rangle\sim \rm{const}$ and $t_{\rm
  out}\sim {\rm const}$ whatever the wind.}. Their maximum  outburst time is obtained
when $\langle\dot{M}_{\rm wind}\rangle = 0$ and the minimum one when
$\langle\dot{M}_{\rm wind}\rangle = \langle\dot{M}_{1}\rangle = \dot{M}_{\rm
Edd}$.

\begin{figure}
\center
\includegraphics[width=9cm] {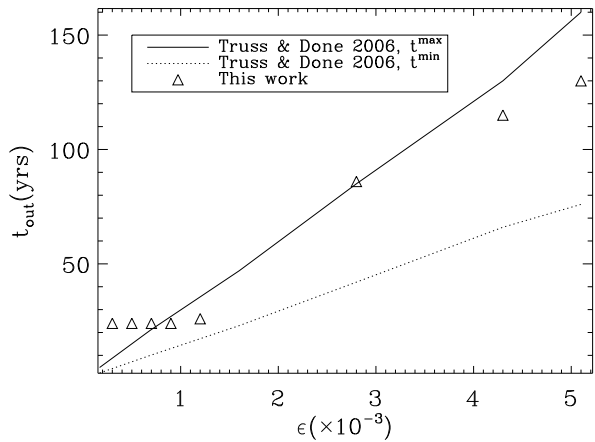}
\caption{Comparison between this work and \protect\citet{Truss06}} 
\label{fig:truss_me_comp}
\end{figure}

In fig.~\ref{fig:truss_me_comp} are plotted the outburst times predicted by
\citet{Truss06} and our work, against the irradiation efficiency.
For values of $\epsilon < 0.7 \times 10^{-3}$, our outburst durations are 
larger than those of \citet{Truss06}. This is due to the fact that 
their model is solely based on the irradiation to switch the disc into the hot
state. Therefore, for small values of $\epsilon$ their predicted outburst 
times vanishes to zero. As discussed in \S\ref{subsec:effstuff} in the low
$\epsilon$ regime of our simulations, the small irradiation radius has the mere effect of keeping 
a small amount of mass in the hot state a bit longer than without irradiation.
When $\epsilon  \gtrsim  10^{-3}$  and the effect of the irradiation
becomes significant, the two approaches  are in fair agreement. 
The duration of the current outburst of \1915 in conjunction with these models
may provide some evidence of the irradiation efficiency in this system.

The black hole LMXB 4U 1755-33 is very reminiscent of \1915.
It went through a $\sim 25$ year long outburst that ended around 1995, has a
long orbital period of 4.4 days\footnote{Long orbital period compared to usual
LMXBs.} and also shows jet-like signatures \citep{2003ApJ...586L..71A, 2005ApJ...618L..45P,
  2006ApJ...641..410K}. To the authors' knowledge, 
the masses of the binary components have not been determined yet, which
exclude any direct attempt to dynamically model this system. However, its outburst
duration lies at the lower end of the range we find \1915 which is 
reasonable given that its orbital period, hence accretion disc, is smaller.

\subsection{Central accretion rate and the need for a wind \label{subsec:discuss_mdot}}
The accretion rate measured at the inner boundary of the disc 
($r_{\rm in}=0.04a \sim 10^{11}$~cm) in the simulations yield $\dot M_{\rm in}\sim 10^{-5}\;\msun$~yr$^{-1}$.
As mentioned in \S\ref{subsec:typ_beh}, this is much larger than the
Eddington rate, at which the black hole is observed to accrete, i.e. $\dot
M_1\sim 10^{-7}\;\msun$~yr$^{-1}$ at $R_{\rm Schw}=4\times 10^{6}$~cm.

So far, the outburst of \1915 has lasted 17 years. Using \eq{eqn:out_time}
with $H/R=0.03$ and $\alpha_h=0.1$ gives a minimum outer radius $R_{\rm out}\sim2\times
10^{12}$~cm. Assuming the surface density profile of the disc follows
$\Sigma_{\rm max}$ (given by \eq{eqn:gradmax}), a rough upper estimate of the mass lost in the
outburst is
\beq
M_{\rm out}=\int_0^{R_{\rm out}}2 \pi r \Sigma_{\rm max}(r) dr \sim 2\times10^{-4}\; \msun.
\eeq
If the outburst were to stop in the near future, this mass implies a transfer rate
through the disc $\dot M_{\rm in}\sim 10^{-5}\;\msun$~yr$^{-1}$, in agreement
with the transfer rate we measure at the inner boundary of the simulations.

Conversely, if one takes the observationnaly inferred Eddington rate 
($\dot M_1\sim10^{-7}\;\msun$~yr$^{-1}$) as the transfer rate through
the disc and use $t_{\rm out}=17$~yrs, one gets $M_{\rm out} \sim
10^{-6}\;\msun$. Assuming the same density profile gives $R_{\rm
  out}\sim 2\times 10^{11}$~cm, which is incompatible with a viscous/outburst
time of 17 years. Hence, it would seem a wind or outflow must carry away the
excess mass and we suggest that most ($\gtrsim 90\%$) of the mass flow through
the disc must be lost in this way.

We note that \citet{2009arXiv0901.1982U} recently concluded in the existence of a
thermally and/or radiation-driven disc wind in \1915 from the absorption 
lines in its spectra and inferred a launching radius $\sim 10^{5}
R_{\rm Schw}$. This corresponds to the innermost regions our simulations.
However, they estimate the mass loss rate in the wind to be 
$\sim 10^{-7}\;\msun$~yr$^{-1}$, of the same order that the accretion rate
onto the black hole. However, such mass loss rates are difficult to
measure and we suggest that the loss rate could be significantly higher.

\section{Conclusion}
In this work, we have studied the long term evolution of the \1915 LMXB 
using global SPH simulations of its accretion disc, which include the 
two-alpha disc instability model, irradiation and wind. The parameter space has
been scanned with the limitation of retaining transient behaviour in the
simulations. We showed that all parameters (but for the wind efficiency) can
play a significant role in the determination of the outburst and recurrence times.

In agreement with both intuition and previous work, the outburst duration is 
affected by $\alpha_h$ (the smaller $\alpha_h$, the longer the outburst) but
not by $\alpha_c$. The opposite effect is noted for the recurrence time (the
smaller $\alpha_c$, the longer the recurrence time and a very marginal effect
of $\alpha_h$.)

Increasing the slope of the critical density profiles is also important: i) the 
outburst is shorter if $\Sigma_{\rm min}$ is increased but only marginally
affected by $\Sigma_{\rm max}$, ii) an increase of  $\Sigma_{\rm min}$ and  
$\Sigma_{\rm max}$ respectively decreases and increases the recurrence time of
the system.

When the irradiation efficiency is small ($\epsilon<10^{-3}$) the
variations of $t_{\rm out}$ and $t_{\rm quiesc}$ are non negligible but span on a relatively narrow
range: $t_{\rm out}\in [20-40]$~years, $t_{\rm quiesc}\in [7000-20000]$~years and 
the duty cycle $DC\in[0.1-0.5]\%$. However, as soon as the irradiation becomes 
non-negligible in determining the outer radius of the hot state region of the
disc, the outburst duration increases considerably, potentially reaching 130
years for $\epsilon=5\times 10^{-3}$.

In \citet{Truss06}, the authors relied on the irradiation 
to set the outburst duration. For large values of the irradiation efficiency,
our results are in agreement with their analytical modelling. However, 
for small values of $\epsilon$, the thermal-viscous instability is at
the origin of the outburst and we show that
\1915's outburst can be expected to last at least $\sim 20 \pm 5 {\rm ~ yr}$.
As the outburst began in 1992 this raises the possibility that the outburst
could end in the next decade. If so this would indicate that the X-ray
irradiation efficiency is small, $\epsilon\lesssim 1.5\times 10^{-3}$. 
If however the outburst persists any longer, the conclusion that significant 
fractions of the outer disc are being irradiated is unavoidable.

The extraordinary duration of the outburst of \1915 (and possibly of 4U
  1755-33) is due to its long orbital period resulting in an unusually large
  amount of gas available in its accretion disk. These extraordinarily long
outbursts of long period, transient LXMBs should not be confused with the
persistent outbursts of short period systems such as Sco-X1
  or 4U 1957+11, in which stable, persistent accretion can be easily established
  by irradiation at moderate mass transfer rates \citep{1996ApJ...464L.127K}.

Except if the irradiation efficiency is high, the duty cycle of \1915 is
$\sim 0.5\%$, at best. Our canonical values gives $DC=0.2\%$: these values are
slightly smaller than the typical duty-cycles measured in shorter periods LMXBs ($\sim
1\%$), or that generally assumed for population models and X-ray luminosity 
functions of nearby galaxies \citep{2004ApJ...601L.147B,Ivanova06}. This
suggests than one needs to be careful when considering long period
systems (for which very little (long-term evolution) data exists) and 
we will consider a more generic approach to these objects in a forthcoming
paper.

\section*{Acknowledgments}
The authors thank the anonymous referee for their comments that have 
helped putting this work into a broader context. CC and GW acknowledge support
from the theoretical astrophysics group STFC rolling at the university of 
Leicester. PD acknowledges a STFC studentship.

\bibliographystyle{mnras} 
\bibliography{deeganetal}

\end{document}